\documentclass[a4paper,twocolumn,showpacs,amsmath,amssymb,pre]{revtex4}
\usepackage{graphicx}
\usepackage{latexsym}
\begin{document}
 
\title[Evidence for a 3-state non-cooperative process]{Coil-helix transition in poly(L-glutamic acid) : Evidence for a 3-state non-cooperative process}

\author{Gilbert Zalczer}
\affiliation{Service de Physique de l'Etat Condens\'{e}, CEA Saclay, 91191 Gif-sur-Yvette cedex, France.}
\email{Gilbert.Zalczer@cea.fr}
 
\begin{abstract}{ A careful analysis of measurements of circular dichroism of poly(L-glutamic acid) (PGA) shows that the data can be very accurately described by introducing a third state for the PGA configuration, in addition to the helix and coil ones, and considering a simple equilibrium between these three states, without cooperativity. The third state is more conspicuous when high molecular weight polyethyleneglycol (PEG) is added. Excluded volume effects shown by differences in presence of short and long PEG chains indicate a direct interaction of  PEG and PGA rather than an osmotic effect. } 
\end{abstract}
\maketitle 

%\date{\today} 

\section{Introduction}
Proteins are important building blocks of life. They have multiple properties due to the
interplay of their four structural levels : primary (sequence of aminoacids), secondary ( alpha helices,
beta sheets, etc.) tertiary ( arrangements of secondary structures in space) and quaternary (
self assembly of supramolecular structures ). The primary structure is fixed by covalent bonds
and is therefore very stable. The other levels of order are governed by much weaker
interactions (hydrogen bonds, dipolar interactions, etc.) and are therefore sensitive to changes
in the environment such as temperature, pH, etc. Understanding the formation of these
structures is an important step in the understanding of their overall properties. Moreover it is strongly suspected that misfoldings are responsible for serious diseases \cite{Chiti2006}.We focus here
on the formation of alpha helices from random coils in a much simplified system : the poly(L-glutamic acid) (PGA). The helical parts are optically active and their population can be easily and
accurately measured by circular dichroism.

This transition is usually analysed in terms of a theory due to Zimm and Bragg \cite{Zimm1959,Grosberg1994}( or variants thereof \cite{Lifson1961,Farago2002,Ghosh2009}). In this theory
an elementary segment of the chain can exist in two states: either as as disordered chain or
as a loop. A free energy  difference between these states comprises  energy and entropy
terms :

\begin{equation}
\Delta F= \Delta E - T \Delta S
\end{equation}

where $\Delta E$ and $\Delta S$ can be considered as independent of the temperature in a first approach.
 (A more complete description \cite{Munoz1995} introduced a linear temperature variation in $\Delta E$ and a logarithmic one in $\Delta S$. The non linearity in the log function in the range from 280K to 370K is too small to be seen so that the equation applies with a different meaning of E and S). We choose the energy unit such that $k_{B}=1$. In the absence of
other interactions, the relative ratio of states would be given by a  Boltzmann factor $\exp (-\Delta F/T)$. In addition, Zimm and Bragg introduced an interaction between neighbouring elements
which facilitates the formation of a new loop beside an existing one and therefore the
transition of the whole chain. The problem is therefore similar to the 1D Ising model and can
be solved.

Many sets of data have been analysed using this model \cite{Scholtz1991,Baldwin1995,Stanley2008,Ghosh2009} and significant cooperation
parameters determined even though a glance at the figures shows a rather poor fit to the
data. A new set of measurements using solutions with added polyethyleneglycol (PEG) and at pH=3.75 have been performed
using state-of-the-art equipment \cite{Koutsioubas2011} and will be the basis of our analysis.
The results are plotted in figure 1.

\begin{figure}
\centering
\includegraphics [bb = 52 600 290 800]{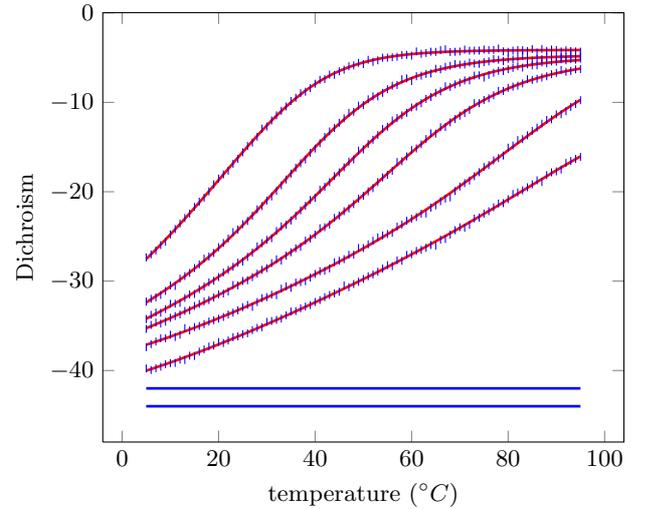}
\caption{Measured dichroism (in $degrees.cm^{2}/mole$) . From top to bottom : $0\%{}$, $5\%{}$, $7.5\%{}$, $10\%{}$,  $20\%{}$ and $30\%{}$ PEG. Red lines are fits using the 3state non-cooperative model. Horizontal lines  show the range of  possible low-temperature asymptots (see text).}
\end{figure}

Without PEG, the transition occurs at low temperature and only
a part of it can be studied. The addition of PEG shifts the transition to higher temperatures but
modifies the system. The figure suggests however that these modifications do not change
qualitatively the basic mechanism.
\section{Critical plot of the data}
A more clear picture can be obtained by plotting the data in a different way : Eq1 can be
rewritten as :
\begin{equation}
\log [x/(1-x)]=\Delta E / T - \Delta S
\end{equation}

where $x$ is the fraction of helices. In the absence of cooperativity a plot of $\log(x/(1-x))$ vs $ 1/T$ would
be a straight line and  strong cooperativity would lead to a sigmoidal shape. 

Computing the helical fraction is, however, not straightforward because it requires the knowledge of the asymptotic values reached by the curves of fig 1 at  high and low temperature, corresponding to fractions of helices of 0\%{} and 100\%{} . These are the optical properties of given molecular conformations and should not vary appreciably with temperature.  The high temperature limit can be safely assessed for the data with 0\%{} and 5\%{} of PEG . A value of $37.4  ^{\circ} $  has been published \cite{Su1994} for the helix maximum contribution (difference between the upper and lower asymptotes). We  allowed for a variation of a few degrees of each of these values to get the most consistent pattern.

\begin{figure}
\centering
\includegraphics[bb=52 600 290 800]{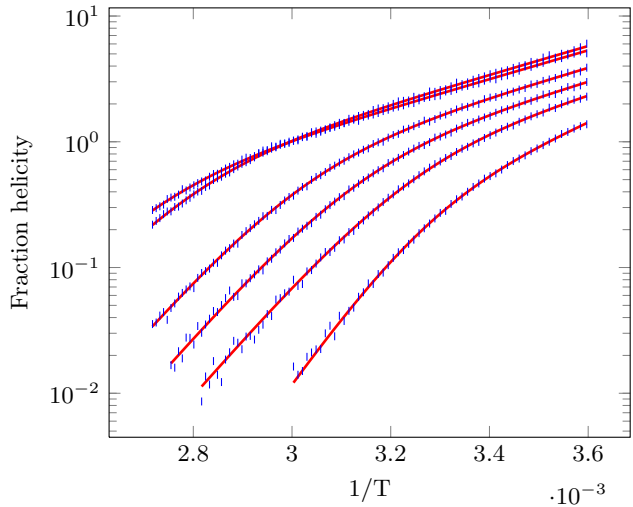}

\caption{Same data as in fig 1 expressed as the ratio of optically active peptide to optically inactive ones vs the inverse temperature. From  bottom to top: $0\%{}$, $5\%{}$, $7.5\%{}$, $10\%{}$,  $20\%{}$ and $30\%{}$ PEG. Red lines are fits using the 3state non-cooperative model. }
\end{figure}
We  notice first that all the curves of fig 2 have a similar shape: two linear parts at low and high temperature matched by a smooth crossover. 
\section{ The 3-state model}
Obviously the data cannot be fitted by any of the shapes predicted by the theory of Zimm and Bragg. However, the presence of two straight
asymptotes hints to the occurence of two transition free energies implying three
thermodynamic states.
The fraction of helices, in the absence of cooperativity, is easily computed from Boltzmann
factors :
\begin{equation}
 \frac{x}{(1-x)}  =  \frac{\exp (-\Delta F_{1}/T)}{  1+\exp (-\Delta F_{2}/T)}
 \end{equation} or 
\begin{equation}
 \frac{x}{(1-x)}  = \frac{\exp -(\Delta E_{1}/T-\Delta S_{1})}{  1+\exp -(\Delta E_{2}/T-\Delta S_{2})}
 \end{equation} 
taking the coil state as the reference one.( The curvature of the lines implies that only one state is optically active.)

 Each curve can be fit very accurately using 6 independent parameters. The values found for the limiting values and the energies are reasonable. However, many of these values cannot be considered as precisely determined. The high helicity side of the PEG poor samples and the low helicity side of the PEG rich samples are not sufficiently present in the experimentally accessible range. Moreover a slight change in a limit value leads to a change in the corresponding energy without altering significantly the quality of the fit. Among the many possible ways of constraining the fit, we choosed to explore a possibility suggested by  the figures namely constant energies ( slopes) for all concentrations. The trial energies are taken  from the best fit of the `best balanced' curve at 10\%{} PEG. They are $10400 K$ and $7780 K$ or $21 kCal.M^{-1}$ and $15.5 kCal.M^{-1}$
The fit of the
experimental data with this formula can be considered  as quite satisfactory for all the samples including PEG as can be seen in fig 1 and fig 2.  The data without PEG however did exhibit a systematic deviation and a different set of values of energies had to be found . The values are $13500K$ and $8870K$ or $27 kCal.M^{-1}$ and $17.8 kCal.M^{-1}$. This fit is the one plotted in fig 1 and fig 2.
This `constant energy' constraint is only a heuristic and fits with constant entropies and adjustable energies are as good.

\section{Effect of PEG chain length}
The results of a similar study have been published a few years ago\cite{Stanley2008}. The main difference is that its authors used short PEG chains while Koutsioubas et al. \cite{Koutsioubas2011} use long PEG chains .  These data (picked from the published figure) can be plotted using the same process (figure 3). 
\begin{figure}
\centering
\includegraphics[bb=52 600 290 800]{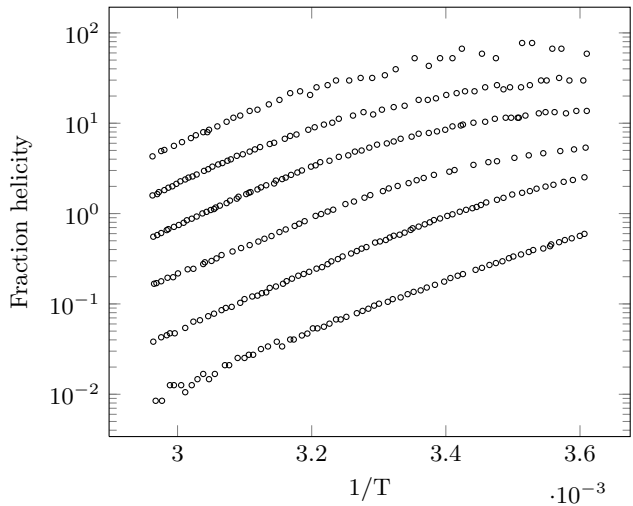}

\caption{Ratio of optically active peptide to optically inactive ones from Stanley and Strey\cite{Stanley2008}. From bottom to top : 10\%{}, 15\%{}, 20\%{}, 25\%{}, 30\%{} and 35 \%{} PEG ( of low molecular weight)}
\end{figure}
A look at figure 3 reveals similarities and differences. First the pattern of two transitions matched by a crossover is kept, but the third state is much less conspicuous. We shall therefore not try to perform any quantitative analysis for this state. The energy difference between  the coil and helix  states (slope of the data at small 1/T) is equal to that observed for long chains.
A striking feature is that the addition of some short PEG has always a roughly constant effect, while the shift of the data due to adding long PEG clearly levels-off. To illustrate this point we have plotted the temperature $T_{1}=E_{1}/S_{1}$ vs. the fraction of PEG for all the curves (figure 5) . The data cannot be compared one to one because of the different pH but the trend of the curves seems significant. The idea is that a peptide can interact with more and more small molecules when the concentration of these molecules is increased, while the excluded volume effects prevent this for macromolecules. This shift has therefore probably an entropic origin. A mechanism related to osmotic stress, as suggested by Stanley and Strey \cite{Stanley2008} should depend on the concentration of PEG but not on the length of the chains.
\begin{figure}
\centering
\includegraphics[bb=52 600 290 800]{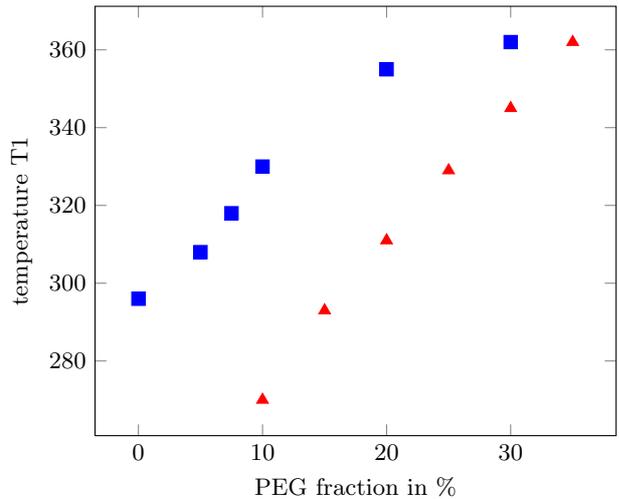}
\caption{Temperature $T1$ at which the ratio $helix/coil=1$ vs PEG concentration. Blue squares : long PEG chains , red triangles : short PEG chains}
\end{figure}
\section{Final remarks}
The presence of a third state betweem coil and helix should not be too surprising. Indeed the
amino acids involved are very prone to interacting with each other leading to different
structures. A hairpin structure, precursor of a beta sheet , could be a likely hypothesis.
A similar  study with different lengths of another peptide
molecules (AEAAKA) has been published \cite{Scholtz1991,Baldwin1995}. Our analysis of the data  from the published figures shows that the
third state is not visible for the shortest chains (up to 26 residues) but seems to appear for the
longer ones (32 or more residues). This clearly favors a third state induced by a self interaction
of the chains.
The characterisation of the third state by neutron or X-ray scattering can be envisionned but
the simultaneous presence of several conformations would make the unraveling of the
spectra difficult. A X-ray study by Muroga et al \cite{Muroga1988}concluded  that the observed data did not fit
with what was expected for a coil when a coil was expected. Indeed the observed spectra
decrease more slowly than expected at large wavevector indicating a more compact state of
the scattering objects. A study of the dynamics of the helix-coil transition for PGA also concluded that this process was not a two-state process.\cite{Sharma2006}
\section{Conclusion}
The first conclusion of this study is that the sharpening of the transition due to first neighbour interaction as predicted by Zimm and Bragg \cite{Zimm1959} is nowhere seen. Instead the consideration of three states in mere chemical equilibrium describes accurately the data. The nature of the third state, which may have a significant importance in health related problems  remains  to be investigated as well as the exact interaction of PGA with PEG. The evidence of excluded volume effects indicates a direct interaction. Circular dichroism measurements for PEG concentrations between $0\%{}$ and $5\%{}$ could be useful for this purpose.

%\pagebreak
%\begin{figure}
%\centering
%\includegraphics[bb=52 600 290 800]{fig0.ps}
%\end{figure}

\end{document}